\documentclass{article}

\usepackage{microtype}
\usepackage{graphicx}

\usepackage{colortbl}
\usepackage{xcolor}
\usepackage{hyperref}

\usepackage[accepted]{icml2025}

\usepackage{amsmath}
\usepackage{amssymb}
\usepackage{mathtools}
\usepackage{amsthm}
\usepackage{xspace}

\usepackage[capitalize,noabbrev]{cleveref}

\theoremstyle{plain}

\theoremstyle{definition}

\theoremstyle{remark}

\usepackage[textsize=tiny]{todonotes}

\icmltitlerunning{\OURS: Learning Code Transformation Rules for Code Editing}
\usepackage{style}

\begin{document}

\twocolumn[
\icmltitle{\OURS: Learning Code Transformation Rules for Code Editing}




\begin{icmlauthorlist}
\icmlauthor{Weichen Li}{uchi}
\icmlauthor{Albert Jan}{columbia}
\icmlauthor{Baishakhi Ray}{columbia}
\icmlauthor{Junfeng Yang}{columbia}
\icmlauthor{Chengzhi Mao}{rutgers}
\icmlauthor{Kexin Pei}{uchi}
\end{icmlauthorlist}

\icmlaffiliation{uchi}{The University of Chicago}
\icmlaffiliation{columbia}{Columbia University}
\icmlaffiliation{rutgers}{Rutgers University}

\icmlcorrespondingauthor{Weichen Li}{weichenli@uchicago.edu}
\icmlcorrespondingauthor{Kexin Pei}{kpei@uchicago.edu}

\icmlkeywords{Machine Learning, ICML}

\vskip 0.3in
]



\printAffiliationsAndNotice{}  

\begin{abstract}
Code editing is a foundational task in software development, where its effectiveness depends on whether it introduces desired code property changes without changing the original code's intended functionality. 
Existing approaches often formulate code editing as an implicit end-to-end task, omitting the fact that code editing procedures inherently consist of discrete and explicit steps. 
Thus, they suffer from suboptimal performance and lack of robustness and generalization. 
We introduce \OURS, a code editing framework that makes the code transformation steps explicit. 
Our key insight is to employ a language model (LM) as an inductive learner to extract code editing rules from the training code pairs as concise meta-rule sets.
Such rule sets will be manifested for each training sample to augment them for finetuning or assist in prompting- and iterative-based code editing.
\OURS outperforms the state-of-the-art by an average of 22.7\% in editing performance and 58.1\% in robustness while achieving 20.2\% higher functional correctness across critical software engineering and security applications, LM models, and editing modes.
\end{abstract}

\section{Introduction}

\begin{figure}[!t]
    \centering
    \includegraphics[width=1\linewidth]{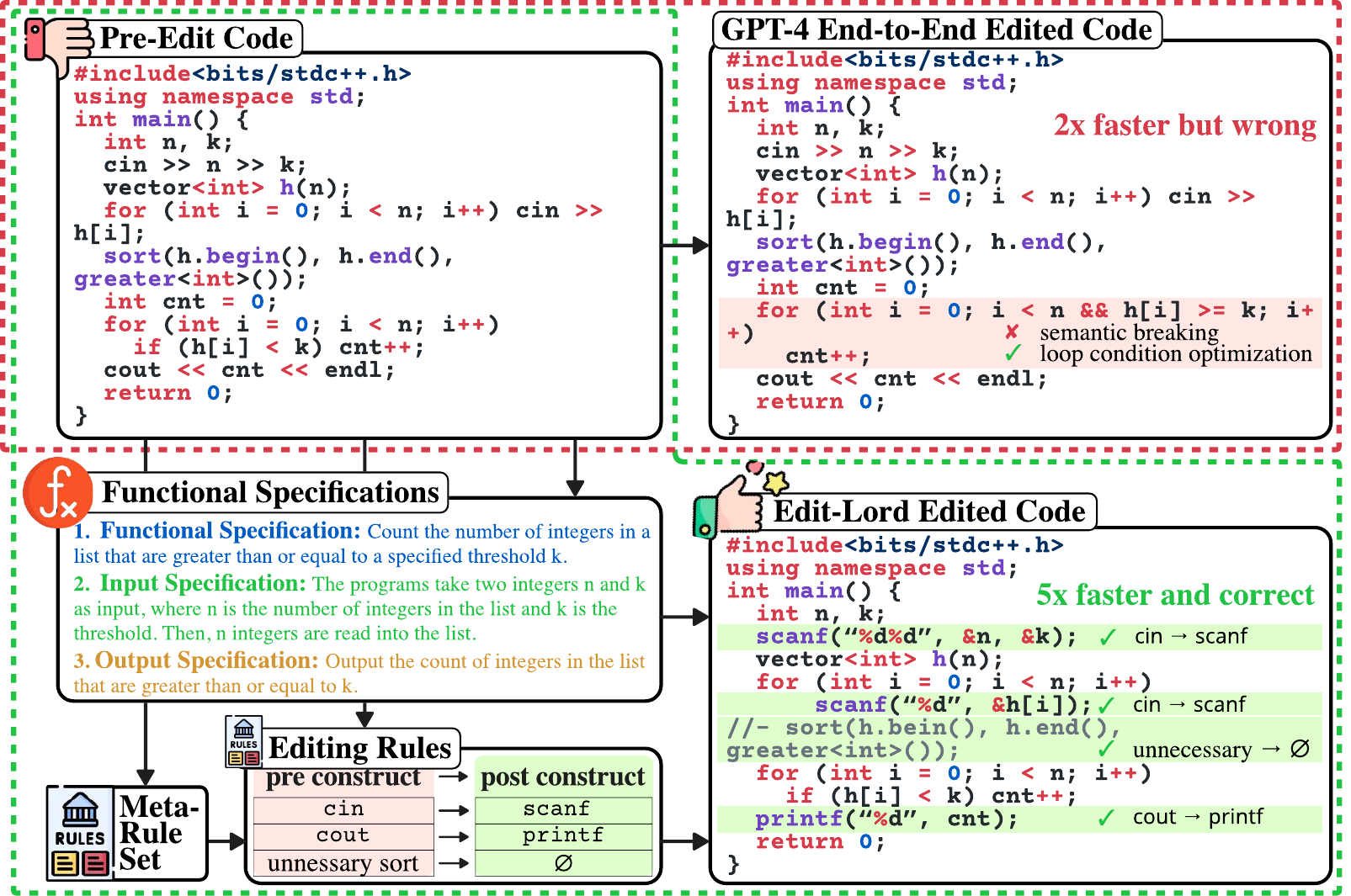}
    \caption{A performance editing example showing how \OURS differs from the existing approach~\cite{pieperf}. The editing rule library here is discovered during the induction rule learning phase (\cref{subsec:method_rule_learning}) and used to prepare the finetuning set or directly prompting (\cref{subsec:method_editing_paradigm}). \OURS's edits include removing redundant \texttt{sort} and calling efficient I/O functions (\texttt{cin} to \texttt{scanf}), while \citet{pieperf} provides a suboptimal performance improvement while breaking the original code's input-output functionality.} 
    \label{fig:overview}
\end{figure}

Pre-trained code Language Models (code LMs) have shown impressive performance in automating software development and substantially improved the developers' productivity in a wide range of programming tasks~\cite{githubAIImpact}. 
Among these tasks, code editing has been a fundamental building block with broad applications, such as optimizing code efficiency~\cite{pieperf, huang2024effi, peng2024perfcodegen, garg2022deepperf}, reverse engineering and decompilation~\cite{tan2024llm4decompile, hu2024degpt, wong2023refining, xie2024resym}, and vulnerability repair and security hardening~\cite{xia2024automated, fu2022vulrepair, peng2025cweval, he2023sven, perry2023users, bhatt2023purple}.

Code editing is often considered more challenging than code generation, as it comes with extra requirements for introducing desired new properties, e.g., improved efficiency, in addition to generating functionally correct code.
Such new requirements necessitate diagnosing and localizing specific properties in the original code, determining the necessary changes to edit these properties, and ensuring the edits do not introduce unintended side effects.
These abstract editing procedures are often compositional because fulfilling certain code property changes can consist of multiple editing steps at multiple program points.
However, they are also modular and reusable, as the same abstract editing steps can be reapplied in diverse code editing scenarios.

Unfortunately, existing works often treat the knowledge of code editing as latent and implicit, and resort to brute-force finetuning to internalize it as part of the model weights~\cite{pieperf, tan2024llm4decompile}.
Such approaches could fail to decouple the modular code transformation procedures from the specific training samples and thus suffer from suboptimal performance (\cref{subsec:eval_main_results}) and the lack of robustness and generalization (\cref{subsec:eval_generalization_robustness}).

We introduce \OURS, a novel framework for learning explicit meta-program transformation rules for code editing tasks. 
Instead of directly prompting or finetuning an LM to perform the editing, \OURS first employs an LM to discover a concise set of inductive transformation rules from training data and then trains an LM to apply these transformations for each code pair. 
\OURS consists of three key subtasks: discovering edit rules to abstract the editing steps, summarizing functional specifications to maintain the code functionality, and learning the discovered rules for code editing. 
\cref{fig:overview} compares \OURS to the finetuned baseline~\cite{pieperf}. 

A key advantage of our approach is that it operates in a structured, discrete transformation space, reflecting the symbolic and compositional nature of code editing tasks.
Since our discovered meta-program transformation rules are expressed in natural language, they remain interpretable. 
Moreover, by exposing the editing steps as explicit modules, \OURS enables more precise operations and allows human experts to intervene when necessary.

We evaluate \OURS on the three critical software engineering and security code editing tasks, including optimizing code efficiency, improving the readability of decompiled code, and repairing security vulnerabilities.
\OURS outperforms the state-of-the-art code editing techniques by 23.3\%, 12.7\%, and 27.6\%, respectively, across multiple code LMs.
With explicit editing rules, \OURS significantly improved the baseline in generalization, e.g., by up to 24.87\% improvement in length generalization and 58.1\% improved robustness against semantics-preserving code transformations. 
In addition to finetuning, \OURS applies to different editing modes, with an average of 56.3\% improvement in zero-shot prompting and 5.7\% improvement for iterative refinement based on execution feedback.
As \OURS exposes the explicit editing steps, we show that its editing performance can be further improved by up to 35.5\% when steered by human experts.
\section{Methodology}
\label{sec:method}
\begin{figure*}[!t]
    \centering
    \includegraphics[width=1.0\linewidth]{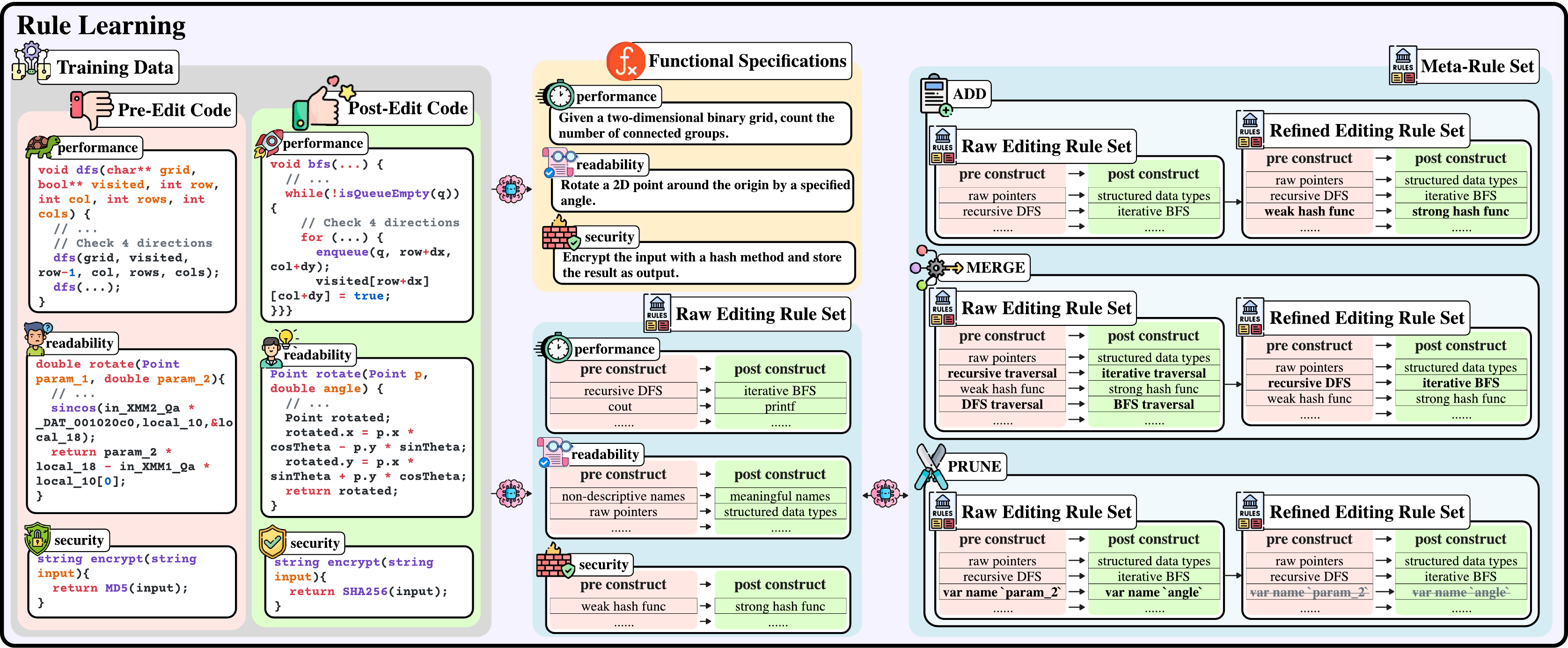}
    \includegraphics[width=1.0\linewidth]{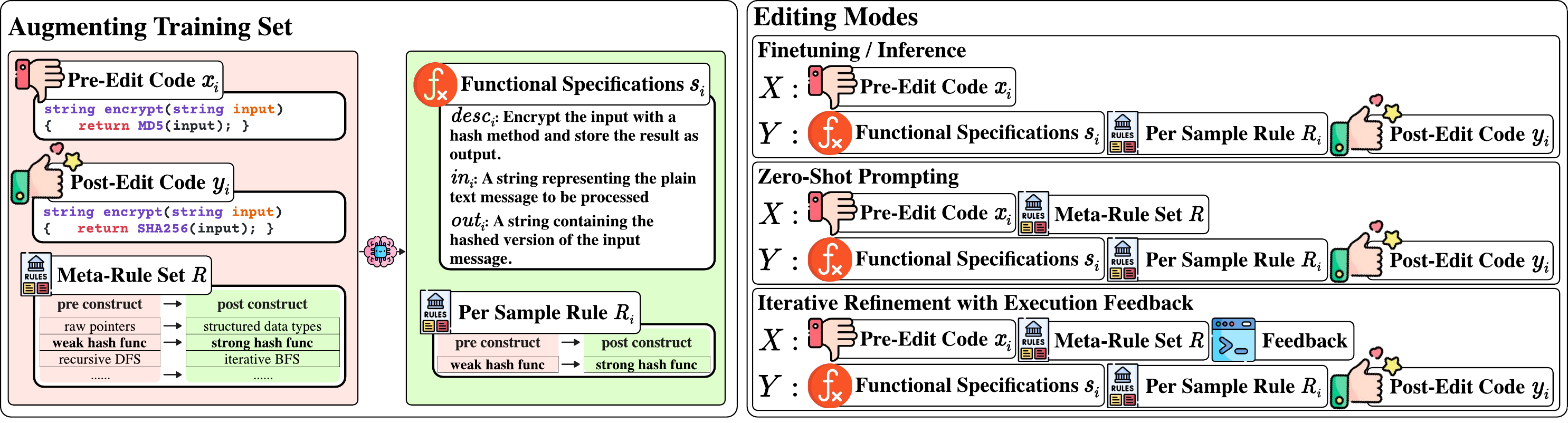}
    \caption{\OURS workflow. The upper section shows how \OURS discovers the meta-rule set and functional specification based on the pre-edit and post-edit training code samples. The lower section shows how \OURS augment the training data. The editing process can then be performed either by querying the finetuned LMs (by augmented training samples) or using the meta-rule set as the prompt to guide the zero-shot prompting and the iterative refinement with external feedback. Note that the rule learning steps for three tasks are \emph{independent}. We put them together just to show that the process is generic for different editing tasks. }
    \label{fig:methodology}
\end{figure*}

\subsection{Problem Statement}
The code editing task aims to transform a given pre-edit code into a post-edit code. 
The post-edit code must be semantically equivalent to the pre-edit code, i.e., preserving its input-output behavior and possessing the desired new code properties, e.g., improved efficiency, readability, or security (fewer vulnerabilities).

More formally, given a Language Model (LM) $M$ and a training dataset $\mathcal{D} = \{(x_i, y_i)\}_{i=1}^n, \mathcal{D}\subseteq X \times Y$, where $X= \{x_i\}_{i=1}^n$ is the set of pre-edit code samples and $Y = \{y_i\}_{i=1}^n$ is the set of post-edit code samples.
Finetuning $M$ for code editing tasks can be formulated as optimizing the conditional probability $P_{M}(Y|X)$ with respect to $M$.

\OURS first creates an augmented training set $\mathcal{D}^\star$ before finetuning the model to perform rule-based code editing. 
Specifically, there are three steps: (1) data-driven inductive rule discovery, (2) functional specification discovery, and (3) rule-based code editing. 
\cref{fig:methodology} illustrates the high-level workflow of \OURS. 
The first two steps bootstrap the training data $\mathcal{D}$, while the third step finetunes the code LM based on the augmented training set $\mathcal{D^\star}$.

In data-driven inductive rule discovery, we query an LM $G$ to iterate the training set and summarize a meta-rule set $R$.
We will elaborate on the process in \cref{subsec:method_rule_learning}.
In functional specification discovery, we extract the functional description $s_i\in S$ that describes the shared input-output behavior of both $x_i$ and $y_i$. 
Such a behavior is expected to be the same because the post-edit code $y_i$ should preserve the same input-output semantics of $x_i$.
With the meta-rule set $R$ and the functional specifications $S$, we obtain an augmented training set $\mathcal{D^\star}\subseteq X\times Y\times R\times S$.

During finetuning, we train the LM $M$ to predict the per-sample editing rules $R_i$, functional specification $s_i$, before predicting the post-edit code $y_i$, all conditioned on the pre-edit code $x_i$: $P_M(y_i|x_i)=\sum_{s_i,R_i}P(s_i|x_i)P(R_i|x_i,s_i)P(y_i|x_i,s_i,R_i)$.
Besides finetuning, we also consider other editing modes, e.g., prompting (\cref{subsec:method_editing_paradigm}).

\subsection{Data-Driven Inductive Rule Discovery}
\label{subsec:method_rule_learning}

Transformations required to achieve the desired changes in code property can be formalized as editing rules.
Intuitively, these rules represent the explicit knowledge of code edits instead of the implicit weight internalized in the model.
Our goal is to learn such an explicit meta-rule set $R$ from the training samples, before learning to perform the code editing guided by these explicit rules for better generalization.
In the following, we describe how \OURS iterates the training set $\mathcal{D}$ to grow $R$ and ensures the rules in $R$ remain modular and reusable.

\paragraph{Inductive meta-rule set initialization}
Formally, we start by extracting raw editing rules by iterating each training code pair $(x_i, y_i)$.
For each $(x_i, y_i)$, we query an LM $G$ to generate a per-sample raw editing rule set $R_i$ that explicitly describes the code property changes required to transform $x_i$ into $y_i$. 
Each raw rule set $R_i$ is thus defined as $R_i=G(x_i, y_i)$. 
We then aggregate $\{R_1,...,R_n\}$ to initialize the raw editing rule set $R = \bigcup_{i=1}^n R_i$.

\paragraph{Iterative meta-rule set refinement}
While the initial $R$ collects the raw editing rules for each sample, 
they are either too generic or too specific to provide actionable guidance for an effective editing. 
For example, they can be as generic as ``check for potential vulnerability'', or as specific as ``switch from \texttt{a} to \texttt{a+1}'' (with very specific variable names). 
Additionally, multiple editing rules may indicate similar transformations, e.g., ``switch from \texttt{cin} to \texttt{scanf}'' and  ``switch from \texttt{std::cin} to \texttt{scanf}''.
Given these challenges, we introduce an iterative meta-rule set refinement algorithm to ensure the ultimate $R$ is concise and effective. 
We start with this initial $R$ and iteratively refine each rule $r\in R$ until $R$ converges.
\cref{fig:methodology} illustrates the procedures. 

To systematically perform the refinement, we define three operations that $G$ can use to update the meta-rule set $R$.
(1) ADD: $R\cup\{\currule\}\rightarrow R$. 
This operation instructs $G$ to directly add the rule $\currule\in R_i$ derived from the training sample $(x_i,y_i)$ to the meta-rule set $R$.
This step is still necessary as MERGE and PRUNE (described below) can be very aggressive and some useful rules have been inadvertently removed.
(2) MERGE: $R\cup\{r_i\oplus r_j\}\setminus\{r_i, r_j\}\rightarrow R$. 
This operation instructs $G$ to merge two rules $r_i, r_j\in R$ by replacing them with an updated rule $r_i\oplus r_j$. 
(3) PRUNE: $R\setminus\{\currule\}\rightarrow R$. This operation instructs $G$ to remove the rule $\currule$ from the meta-rule set $R$.

\cref{alg:methodology-rule-learning} shows the procedures of updating the editing rule set $R$.
Before adding each $\currule$ (ADD and MERGE) to $R$, we prompt $G$ to assess whether it is a balanced rule, i.e., neither too generic nor too specific (line 4). 
If not, we apply PRUNE to discard it (line 14). 
Otherwise, we prompt $G$ again to decide whether we should ADD or MERGE $\currule$ (line 5).
Specifically, $G$ will be prompted to decide whether there exists a rule in $R$ similar or identical to $\currule$. 
We apply MERGE (line 10) if yes and ADD (line 7) if not.
\cref{fig:methodology} and \cref{tab:rule-example} include some examples of the learned rules in $R$ after this algorithm completes.

\begin{algorithm}[!t]
    \caption{Iterative Meta-Rule Set Refinement}
    \label{alg:methodology-rule-learning}
\begin{algorithmic}[1]
    \INPUT  Initial Meta-Rule Set $R$
    \OUTPUT Finalized Meta-Rule Set $R'$
    \STATE $R'\leftarrow R$
    \WHILE {$R'$ not converge}
        \FOR{\textbf{each} $\currule \in R \cup R'$}
            \IF {$\currule$ is well-balanced}
                \IF{$\nexists r^{*}\in R'$ similar to $r$}
                    \STATE /* Add Rules */
                    \STATE $R' \gets R' \cup \{\,\currule\,\}$
                \ELSE
                    \STATE /* Merge Rules */
                    \STATE $R' \gets R' \cup \{\currule \oplus r^{*}\} \setminus \{\currule, r^{*}\}$
                \ENDIF
            \ELSE
                \STATE /* Prune Rules */
                \STATE $R' \gets R' \setminus \{\,\currule\,\}$
            \ENDIF
        \ENDFOR
    \ENDWHILE
    \STATE \textbf{return} $R'$
\end{algorithmic}
\end{algorithm}

\subsection{Functional Specification Discovery}
We consider functional specification as a high-level natural language description of the program's intended behavior and input-output constraints. 
Specifically, for each sample $(x_i, y_i)$ in the training dataset, we define the functional specification $s_i$ of the training code pair ($x_i, y_i$) as $s_i=\{desc_i, in_i, out_i\}$, where $desc_i$ describes the functionality implemented in ($x_i, y_i$), and $in_i$ and $out_i$ specifies the input and expected output constraints of ($x_i, y_i$), respectively.
We prompt $G$ to generate $s_i$.

\subsection{Code Editing Modes}
\label{subsec:method_editing_paradigm}
\label{subsec:method_finetuning_mode}
\paragraph{Finetuning for rule-based code editing}
Once we obtain the meta-rule set $R$, we iterate each training sample $(x_i, y_i)$, and prompt $G$ to identify $R_i\subseteq R$ as the per-sample editing rules that transform $x_i$ to $y_i$. 
With the per-sample rules $R_i$, we also incorporate the corresponding functional specifications $s_i$ to prepare the augmented finetuning set $D^{\star}=\{(x_i, s_i, R_i, y_i)\}_{i=1}^n$ (as shown in bottom-left in \cref{fig:methodology}). 

The finetuning task can now be formulated as modeling the following conditional probability for each sample: $P_M(y_i, s_i, R_i|x_i)=P(s_i|x_i)P(R_i|x_i,s_i)P(y_i|x_i,s_i,R_i)$.
Note that the LM $M$ to be finetuned may differ from $G$, which is used to construct the meta-rule set $R$ and functional specifications $S$ (see \cref{sec:eval}). 

\paragraph{Other editing modes}
Beyond editing the code using the finetuned model, the generic design of \OURS also supports other editing modes.
For example, with the meta-rule set $R$, we can simply prompt an LM $M$ to directly generate the edited code or further refine the edited code based on external feedback (e.g., execution information~\cite{peng2024perfcodegen}).
\cref{fig:methodology} illustrates other code editing modes, i.e., zero-shot prompting and iterative refinement with execution feedback.
We study how \OURS assists these additional editing modes in \cref{subsec:eval_other_editing_modes}.
\section{Experiments}
\label{sec:eval}

We evaluate \OURS on three critical software engineering and security applications that can be formulated as code editing tasks (\cref{subsec:eval_setup}). 
We consider inference using the finetuned model as our default editing mode and compare it to the state-of-the-art baselines, which are also mostly based on finetuning~\cite{pieperf, tan2024llm4decompile, fu2022vulrepair}.
In \cref{subsec:eval_other_editing_modes}, we also show \OURS complements other editing modes.
We choose models that can be full-parameter finetuned on our local hardware (2x4 Nvidia L40S GPUs), i.e., open-source DeepSeek-Coder 1.3B and 6.7B, or via an online API, i.e., GPT-4o mini.
To generate the functional specifications and editing rules, we use GPT-4o mini ($G$ in \cref{sec:method}). 

\subsection{Setup: Tasks, Datasets, and Metrics}
\label{subsec:eval_setup}

\paragraph{Performance optimization} 
This task aims to edit a given program to improve its execution efficiency (see prompts in \cref{app_sec:prompting-perf}). 
Following \citet{pieperf}, we specifically focus on the execution time speedup and use gem5 CPU simulator ~\cite{binkert2011gem5} to mitigate the noises introduced by the bare metal.
We use the \texttt{HQ} (high quality) dataset from \citet{pieperf} for training and evaluation.
The dataset consists of 4,085 training (slow and fast code) pairs, 2,544 validation samples, and 978 testing samples.

We adopt the same metrics as \citet{pieperf}:
\emph{Correct@$k$}: the percentage of problems in the testing set for which the LM generates at least one correct solution out of the $k$ candidates.
\emph{OPT@$k$}: the percentage of problems in the testing set that the fastest and correct code among the $k$ LM generated programs is at least 10\% faster. 
\emph{Speedup@$k$}: the average absolute ratio between the execution time required by the given slow code and the fastest and correct code among the $k$ LM-generated programs.

\begin{figure}[!t]
    \centering
    \includegraphics[width=1\linewidth]{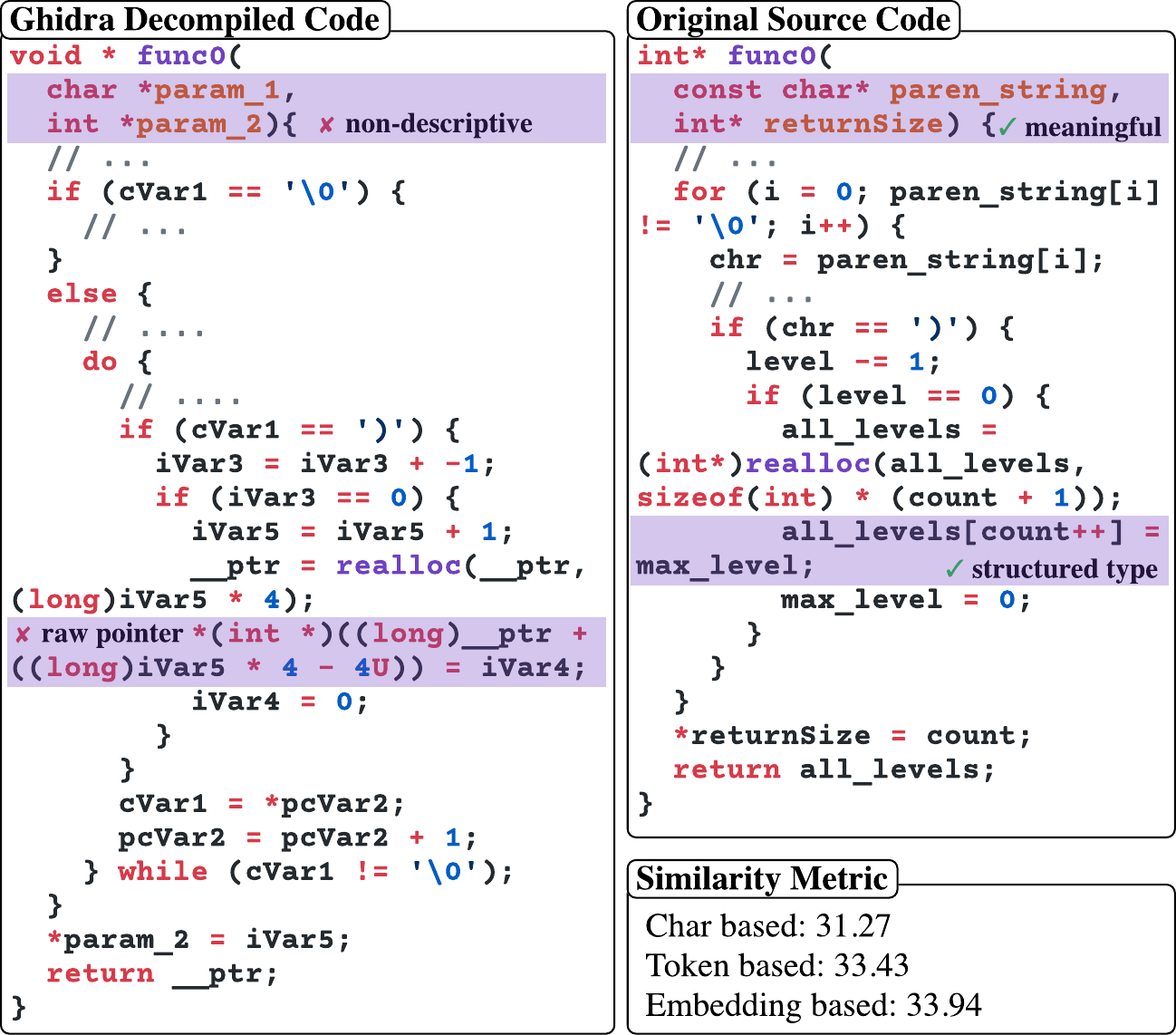}
    \caption{The similarity metrics between the Ghidra decompiled code and its corresponding source code are consistently low. This outcome is primarily due to the non-descriptive variable names and convoluted data structures, which result in both poor readability and consistently low similarity scores.}
    \label{fig:decompile-metric-insight}
\end{figure}

\paragraph{Decompilation} 
This task aims to edit a highly unreadable decompiled code into a more readable form (see prompts in \cref{app_sec:prompting-decompile}).
We obtain these decompiled code samples using the off-the-shelf decompiler Ghidra~\cite{ghidra}.
\cref{fig:decompile-metric-insight} shows an example of the unreadable decompiled code and its original source code.

To construct training and validation samples, we follow \citet{tan2024llm4decompile} to randomly sample original code snippets from AnghaBench~\cite{da2021anghabench} and construct the corresponding Ghidra decompiled code samples by compiling the original code and decompiling it. 
We keep the testing samples strictly non-overlapping with the training by using HumanEval-Decompile~\cite{tan2024llm4decompile}.

Ghidra decompiled code in HumanEval-Decompile may contain both syntactic and semantic errors, i.e., the pre-edit decompiled code may not compile or fail the test cases.
These samples make it impossible for any code editor to reconstruct the semantically correct code without an oracle, which is often not available in typical code editing scenarios (\cref{sec:discussion}).
Therefore, we only consider the functionally correct decompiled code from the HumanEval-Decompile~\cite{tan2024llm4decompile}. 
Overall, our dataset set consists of 8,567 (machine-decompiled code, original source code) training code pairs, 834 validation samples, and 131 testing samples with test cases. 

We adopt the same metrics in \citet{tan2024llm4decompile} and extend its readability measurements. 
\emph{Compilability}: the percentage of problems in the testing set where the model generates the compilable program. 
\emph{Correctness}: the percentage of problems in the testing set where the model generates the correct program.
\emph{Readability}: similarity between the ground truth and the edited code at three different levels, i.e., character-, token-, and embedding-level.
The similarity is defined as $1-d$, where $d$ is the edit distance between the ground truth source code and the recovered original code at the character and token level, and cosine distance between CodeSage~\cite{zhang2024codesage} embeddings of them at the embedding-level.

\paragraph{Security hardening} 
This task aims to edit a given vulnerable code into the patched version (see prompts in \cref{app_sec:prompting-sec}).
In the vulnerable code, there can be one or more vulnerabilities under different Common Weakness Enumeration (CWE) categories.

We obtain the vulnerable and secure code pairs from SVEN~\cite{he2023sven} for training and validation, and evaluate \OURS on a strictly unseen testing set, CWEval~\cite{peng2025cweval}.
CWEval includes vulnerable code samples covering 31 CWEs.
Importantly, we choose CWEval because each of its samples includes both functionality and security tests to automate the evaluation, while the other benchmarks, e.g.,~\cite{fu2022vulrepair} focus only on security fixes and rely on manual effort to check the functionality.

Following CWEval~\cite{peng2025cweval}, we define $n$ as the total number of sampled solutions and a varying $k\leq n$.
We then evaluate the security repair performance with the following three metrics.
\emph{Correct@$k$}: the expectation of any of $k$ LM generated solutions is correct.
\emph{Security@$k$}: the expectation of any of $k$  LM generated solutions is secure. 
\emph{Correct \& Security@$k$}: the expectation of any of $k$  LM generated solutions is both correct and secure. 

These metrics are calculated through the same formula $\mathbb{E}_{\text{Problems}} \left[ 1 - \frac{\binom{n-c}{k}}{\binom{n}{k}} \right]$, where $c$ is the number of correct programs, secure programs, and both correct and secure programs, respectively.

\subsection{Main Results}
\label{subsec:eval_main_results}

\begin{table*}[!t]    
    \centering
    \setlength{\tabcolsep}{2.8pt}
    \renewcommand{\arraystretch}{1.1}
    \small
    
    \caption{Main results for performance optimization, decompilation, and security hardening. ``Finetuned'' refers to the approach from \citet{pieperf}, \citet{tan2024llm4decompile}, \citet{fu2022vulrepair}, respectively.}
    
    \label{tab:main-result}
    \begin{tabular}{rllllll}
    \toprule
    \multirow{2}{*}{Performance opt.}&\multicolumn{2}{c}{OPT@$k\uparrow$} &\multicolumn{2}{c}{Speedup@$k\uparrow$} &\multicolumn{2}{c}{Correct@$k\uparrow$} \\
    \cmidrule{2-7}
     & $k=$1 & $k=$8 & $k=$1 & $k=$8 & $k=$1 & $k=$8 \\
     \midrule
    \multicolumn{7}{l}{DeepSeek-Coder 1.3B} \\ \midrule
    
     Prompt & 4.0 & 16.2 & 1.0 & 1.2 & 11.3 & 18.3\\
     CoT & 5.6 & 15.9 & 1.0 & 1.2 & 10.0 & 19.3\\
     Finetuned & 12.5 & 47.5 & 1.5 & 3.0 & 18.5 & 67.5\\
     \rowcolor{gray!30} \OURS (Ours)  & \textbf{16.4} & \textbf{53.1} & \textbf{1.7} & \textbf{3.8} & \textbf{28.9} & \textbf{72.0}\\  
     \midrule
    \multicolumn{7}{l}{DeepSeek-Coder 6.7B} \\ \midrule
     Prompt & 6.0 & 19.5 & 1.1 & 1.2 & \textbf{60.0} & 89.5\\
     CoT & 4.1 & 6.8 & 1.1 & 1.2 & 16.0 & 36.6\\
     Finetuned  & 7.8 & 76.9 & 2.1 & 3.1 & 20.7 & 87.4\\
     \rowcolor{gray!30} \OURS (Ours) & \textbf{11.0} & \textbf{82.5} & \textbf{2.7} & \textbf{4.5} & 36.0 & \textbf{89.5}\\
     \midrule
    \multicolumn{7}{l}{GPT-4o mini} \\ \midrule
     Prompt & 12.2 & 26.4 & 1.1 & 1.3 & 47.4 & 71.5  \\
     CoT & 10.0 & 29.1 & 1.1 & 1.5 & 45.2 & 64.1\\
     Finetuned  & 28.1 & 61.5 & 2.3 & 3.8 & 59.0 & 89.5\\
     \rowcolor{gray!30} \OURS (Ours)  & \textbf{31.2} & \textbf{72.5} & \textbf{2.9} & \textbf{4.2} & \textbf{62.5} & \textbf{93.5} \\
    \bottomrule
    \end{tabular}
    \setlength{\tabcolsep}{3.5pt}
    \renewcommand{\arraystretch}{1.1}
    \small
    \label{tab:main-result-decompilation}
    \begin{tabular}{rlllll}
    \toprule
    \multirow{2}{*}{Decompilation}& \multirow{2}{*}{Compile$_\uparrow$} & \multirow{2}{*}{Correct$_\uparrow$} & \multicolumn{3}{c}{Readability$_\uparrow$} \\
    \cmidrule{4-6}
    & & & char & token & emb \\
    \midrule
    \multicolumn{6}{l}{DeepSeek-Coder 1.3B} \\
    \midrule
     Prompt & 67.9 & \textbf{65.7} & 31.4 & 35.7 & 31.3\\
     CoT & 10.7 & 6.9 & 31.7 & 35.9 & 31.2\\
     Finetuned   & 77.1  & 38.9 & 36.6 & 40.8 & 37.5 \\
     \rowcolor{gray!30} \OURS  (Ours)     & \textbf{93.1} & 46.6 & \textbf{44.0} & \textbf{47.6} & \textbf{41.4}\\
    \midrule
     \multicolumn{6}{l}{DeepSeek-Coder 6.7B} \\
     \midrule
     Prompt & 83.2 & \textbf{64.9} & 35.3 & 39.3 & 34.3\\
     CoT & 4.4 & 4.4 & 30.8 & 35.4 & 32.7\\
     Finetuned & 89.6 & 56.5 & 42.1 & 47.8 & 39.7\\
     \rowcolor{gray!30} \OURS  (Ours)   & \textbf{90.1} & 58.8 & \textbf{46.2} & \textbf{49.9} & \textbf{43.3}\\
     \midrule
     \multicolumn{6}{l}{GPT-4o mini} \\
     \midrule
     Prompt & 61.8 & 44.3 & 33.1 & 37.0 & 37.9\\
     CoT & 59.5 & 46.6  & 34.3 & 38.3 & 39.8\\
     Finetuned   & 86.3 & 52.7 & 45.6 & 49.4 & 43.6\\
     \rowcolor{gray!30} \OURS  (Ours)    & \textbf{97.7} & \textbf{67.2} & \textbf{51.4} & \textbf{54.9} & \textbf{48.4}  \\
    \bottomrule
    \end{tabular}
    \centering
    \setlength{\tabcolsep}{5pt}
    \renewcommand{\arraystretch}{1.1}
    \small
    \label{tab:main-result-security}
    \begin{tabular}{rlllllllll}
        \toprule
        \multirow{2}{*}{Security hardening}&\multicolumn{3}{c}{Correct@$k\uparrow$} &\multicolumn{3}{c}{Security@$k\uparrow$} &\multicolumn{3}{c}{Correct \& Sec@$k\uparrow$} \\
        \cmidrule{2-10}
        & $k=1$ & $k=10$ & $k=50$ & $k=1$ & $k=10$ & $k=50$ & $k=1$ & $k=10$ & $k=50$ \\
        \midrule
        \multicolumn{10}{l}{DeepSeek-Coder 1.3B} \\
        \midrule
        Prompt & 6.7 & 26.8 & 40.4 & 3.0 & 15.2 & 30.8 & 1.0 & 7.1 & 15.4\\
        CoT & 21.2$_{+14.5}$  & 39.7$_{+12.9}$  & 48.1$_{+7.7}$  & 6.4$_{+3.4}$  & 20.2$_{+5.0}$  & 34.6$_{+3.8}$  & 1.8$_{+0.8}$  & 9.4$_{+2.3}$  & 17.3$_{+1.9}$ \\
        Finetuned & 24.1$_{+17.4}$  & 35.7$_{+8.9}$  & 40.4$_{+0.0}$  & 9.5$_{+6.6}$  & 21.5$_{+6.3}$  & 28.9$_{-1.9}$  & 5.3$_{+4.3}$  & 13.8$_{+6.8}$  & 19.2$_{+3.8}$ \\
        \rowcolor{gray!30} \OURS  (Ours)  & \textbf{31.2$_{+24.5}$}  & \textbf{49.6$_{+22.8}$}  & \textbf{59.6$_{+19.2}$}  & \textbf{12.2$_{+9.2}$}  & \textbf{28.4$_{+13.2}$}  & \textbf{36.5$_{+5.8}$}  & \textbf{7.4$_{+6.4}$}  & \textbf{18.7$_{+11.6}$}  & \textbf{23.1$_{+7.7}$} \\
        \midrule
        \multicolumn{10}{l}{DeepSeek-Coder 6.7B} \\
        \midrule
        Prompt & 20.3 & 31.7 & 44.2 & 14.8 & 29.5 & 36.5 & 8.7 & 17.3 & 23.1\\
        CoT & 13.0$_{-7.3}$  & 40.3$_{+8.6}$  & 51.9$_{+7.7}$  & 7.6$_{-7.2}$  & 32.7$_{+3.2}$  & 42.3$_{+5.8}$  & 8.8$_{+0.0}$  & 17.8$_{+0.5}$  & 21.1$_{-1.9}$ \\
        Finetuned & 22.4$_{+2.1}$  & \textbf{45.5$_{+13.8}$}  & \textbf{53.9$_{+9.6}$}  & 13.7$_{-1.1}$  & 35.0$_{+5.4}$  & 44.2$_{+7.7}$  & 7.7$_{-1.0}$  & 20.6$_{+3.3}$  & 25.0$_{+1.9}$ \\
        \rowcolor{gray!30} \OURS  (Ours)   & \textbf{25.3$_{+5.0}$}  & 41.3$_{+9.6}$  & 49.1$_{+4.9}$  & \textbf{17.9$_{+3.1}$}  & \textbf{37.1$_{+7.6}$}  & \textbf{50.0$_{+13.5}$}  & \textbf{11.8$_{+3.1}$}  & \textbf{23.9$_{+6.6}$}  & \textbf{30.8$_{+7.7}$} \\
        \midrule
        \multicolumn{10}{l}{GPT-4o mini} \\
        \midrule
    Prompt & 21.9 & 36.2 & 46.1 & 18.0 & 30.2 & 40.4 & 13.0 & 23.1 & 26.9\\
    CoT & 27.4$_{+5.5}$  & 43.2$_{+7.0}$  & 53.9$_{+7.7}$  & 23.3$_{+5.2}$  & 35.3$_{+5.1}$  & 42.3$_{+1.9}$  & 18.4$_{+5.4}$  & 28.1$_{+5.1}$  & 34.6$_{+7.7}$ \\
    Finetuned & 30.1$_{+8.3}$  & 42.4$_{+6.2}$  & 50.0$_{+3.9}$  & 16.5$_{-1.6}$  & 28.5$_{-1.7}$  & 34.6$_{-5.8}$  & 13.0$_{+0.0}$  & 22.1$_{-1.0}$  & 30.8$_{+3.8}$ \\
    \rowcolor{gray!30} \OURS  (Ours)   & \textbf{31.4$_{+9.5}$}  & \textbf{49.9$_{+13.6}$}  & \textbf{61.5$_{+15.4}$}  & \textbf{27.6$_{+9.5}$}  & \textbf{40.3$_{+10.1}$}  & \textbf{50.0$_{+9.6}$}  & \textbf{21.2$_{+8.2}$}  & \textbf{29.3$_{+6.2}$}  & \textbf{36.5$_{+9.6}$} \\
        \bottomrule
    \end{tabular}
\end{table*}

We compare \OURS to zero-shot prompting, chain-of-thought prompting (CoT), and finetuning (state-of-the-art baselines) on three tasks: performance optimization, decompilation, and security hardening. 
As illustrated in \cref{tab:main-result}, \OURS outperforms the state-of-the-art baselines across all tasks and models, with 23.3\%, 12.7\%, and 27.6\% improvements on average in performance optimization,  decompilation, and security hardening, respectively. 

While \OURS underperforms the prompting-based approach based on DeepSeek-Coder 1.3B and 6.7B in the decompilation task and DeepSeek-Coder 6.7B in the performance task on the functional correctness, it is important to note that this metric alone can often appear overstated by simply keeping pre-edit code unmodified. 
The readability metrics demonstrate that the prompting approach even \emph{decreases} the original Ghidra decompiled code's readability, i.e., 36, 31.7, and 31.2 for the character-, token- and embedding-level readability metrics, respectively (\cref{subsec:eval_setup}).

\subsection{Ablations}
\label{subsec:eval_ablations}
\begin{table*}[!t]

    \centering
    \setlength{\tabcolsep}{2.8pt}
    \renewcommand{\arraystretch}{1.1}
    \small

    \caption{Ablations on performance optimization, decompilation, and security hardening (based on DeepSeek-Coder 1.3B).}
    \label{tab:ablation}
    
    \begin{tabular}{rllllll}
    \toprule
    \multirow{2}{*}{Performance opt.}&\multicolumn{2}{c}{OPT@$k\uparrow$} &\multicolumn{2}{c}{Speedup@$k\uparrow$} &\multicolumn{2}{c}{Correct@$k\uparrow$} \\
    \cmidrule{2-7}
     & $k=$1 & $k=$8 & $k=$1 & $k=$8 & $k=$1 & $k=$8 \\
     \midrule
     Finetuned & 12.5 & 47.5 & 1.5 & 3.0 & 18.5 & 67.5\\
     + func-spec & 12.5 & 49.2 & 1.7 & 3.5 & 21.4 &  \textbf{73.4}\\
     + edit-rule & 15.7 & 51.9 &  1.8 & 3.7 & 26.2 & 70.0\\
     \rowcolor{gray!30} \OURS  (Ours)   & \textbf{16.4} &  \textbf{53.1} &  \textbf{1.8} &  \textbf{3.8} &  \textbf{28.9} & 72.0\\ 
    \bottomrule
    \end{tabular}
    \setlength{\tabcolsep}{3.5pt}
    \renewcommand{\arraystretch}{1.1}
    \small
    \begin{tabular}{rlllll}
    \toprule
    \multirow{2}{*}{Decompilation}& \multirow{2}{*}{Compile$_\uparrow$} & \multirow{2}{*}{Correct$_\uparrow$} & \multicolumn{3}{c}{Readability$_\uparrow$} \\
    \cmidrule{4-6}
    & & & char & token & emb \\ 
    \midrule
     Finetuned   & 77.1 & 38.9 & 36.6 & 40.8 & 37.5 \\ 
     + func-spec & \textbf{93.9} & 38.9 & 38.9 & 43.6 & 38.4\\
     + edit-rule & 87.0 & 29.8 & 41.5 & 45.2 & 39.5\\
     \rowcolor{gray!30} \OURS  (Ours)       & 93.1 & \textbf{46.6} & \textbf{44.0} & \textbf{47.6} & \textbf{41.4}\\
    \bottomrule
    \end{tabular}
    \setlength{\tabcolsep}{5.7pt}
    \renewcommand{\arraystretch}{1.1}
    \small
    \begin{tabular}{rlllllllll}
    \toprule
    \multirow{2}{*}{Security hardening}&\multicolumn{3}{c}{Correct@$k\uparrow$} &\multicolumn{3}{c}{Security@$k\uparrow$} &\multicolumn{3}{c}{Correct and Sec@$k\uparrow$} \\
    \cmidrule{2-10}
    & $k=1$ & $k=10$ & $k=50$ & $k=1$ & $k=10$ & $k=50$ & $k=1$ & $k=10$ & $k=50$ \\
    \cmidrule{2-10}
Finetuned & 24.1 & 35.7 & 40.4 & 9.5 & 21.5 & 28.9 & 5.3 & 13.8 & 19.2\\
+func-spec & 25.5$_{+1.3}$  & 41.8$_{+6.1}$  & 46.1$_{+5.8}$  & 11.5$_{+1.9}$  & 27.3$_{+5.8}$  & 36.5$_{+7.7}$  & 6.0$_{+0.7}$  & 15.8$_{+2.0}$  & 21.1$_{+1.9}$ \\
+edit-rule & 25.7$_{+1.6}$  & 39.0$_{+3.3}$  & 44.2$_{+3.8}$  & 11.8$_{+2.3}$  & 27.3$_{+5.8}$  & \textbf{38.5$_{+9.6}$ } & 6.7$_{+1.3}$  & 17.5$_{+3.6}$  & 23.1$_{+3.8}$ \\
\rowcolor{gray!30} \OURS  (Ours)   & \textbf{31.2$_{+7.1}$}  & \textbf{49.6$_{+13.9}$}  & \textbf{59.6$_{+19.2}$}  & \textbf{12.2$_{+2.6}$}  & \textbf{28.4$_{+6.9}$}& 36.5$_{+7.7}$  & \textbf{7.4$_{+2.1}$}  & \textbf{18.7$_{+4.8}$}  & \textbf{23.1$_{+3.8}$} \\

     \bottomrule
    \end{tabular}
\end{table*}
\begin{table}[!t]

    \centering
    \setlength{\tabcolsep}{3pt}
    \renewcommand{\arraystretch}{1.1}
    \small
    
    \caption{Comparing \OURS to \citet{tan2024llm4decompile} against semantics-preserving code transformations for robustness and unseen samples with longer lengths for generalization by measuring performance degradation with the original decompilation results (\cref{tab:main-result-decompilation}).}
    \label{tab:robustness-decompile-obf}
    \label{tab:generalization-decompile-length}
    \begin{tabular}{rlllll}
    \toprule
    & \multirow{2}{*}{Compile$_\uparrow$} & \multirow{2}{*}{Correct$_\uparrow$} & \multicolumn{3}{c}{Readability$_\uparrow$} \\
    \cmidrule{4-6}
    & & & char & token & emb \\
    \midrule 
    \multicolumn{6}{l}{Robustness}\\
    \midrule
    Finetuned & 7.0 & 8.6 & 4.2 & 3.1 & 5.5 \\
    \rowcolor{gray!30} \OURS (Ours)   & \textbf{1.2} & \textbf{4.5} & \textbf{1.8} & \textbf{1.7} & \textbf{4.3} \\

    \midrule
    \multicolumn{6}{l}{Generalization} \\
    \midrule
Finetuned & 4.3 & 11.4 & 7.6 & 8.9 & 5.8 \\
\rowcolor{gray!30} \OURS (Ours) & \textbf{3.4} & \textbf{9.5} & \textbf{5.8} & \textbf{8.7} & \textbf{4.3} \\

    \bottomrule
    \end{tabular}
\end{table}

\begin{table*}[!t]

    \centering
    \setlength{\tabcolsep}{3.5pt}
    \renewcommand{\arraystretch}{1.1}
    \small
    
    \caption{Editing rules augmented by human experts using DeepSeek-Coder 1.3B.}
    \label{tab:human}
    
    \begin{tabular}{rllllll}
    \toprule    
    \multirow{2}{*}{Performance opt.} &\multicolumn{2}{c}{OPT@$k\uparrow$} &\multicolumn{2}{c}{Speedup @$k\uparrow$} &\multicolumn{2}{c}{Correct@$k\uparrow$} \\
    \cmidrule{2-7}
    & $k=1$ & $k=8$ & $k=1$ & $k=8$ & $k=1$ & $k=8$ \\
    \midrule
     \OURS   & 16.4 & 53.1 & 1.7 & 3.8 & 28.9 & 72.0\\  
    \rowcolor{gray!30} + human & \textbf{17.9} & \textbf{61.4} & \textbf{1.9} & \textbf{3.9} & \textbf{53.0} & \textbf{76.9}\\
    \bottomrule
    \end{tabular}
    \hspace{8pt}
    \begin{tabular}{rlllll}
    \toprule
    \multirow{2}{*}{Decompilation} & \multirow{2}{*}{Correct$_\uparrow$} & \multirow{2}{*}{Compile$_\uparrow$} &\multicolumn{3}{c}{Readability$_\uparrow$} \\
    \cmidrule{4-6}
    & & & char & token & emb \\
    \midrule
     \OURS       & 46.6 & \textbf{93.1} & 44.0 & 47.6 & 41.4\\
    \rowcolor{gray!30} + human & \textbf{49.9} & 89.3 & \textbf{47.0} & \textbf{51.3} & \textbf{56.1}\\
    \bottomrule
    \end{tabular}    
    \setlength{\tabcolsep}{6pt}
    \begin{tabular}{rlllllllll}
    \toprule
    \multirow{2}{*}{Security hardening} &\multicolumn{3}{c}{Correct@$k\uparrow$} &\multicolumn{3}{c}{Security@$k\uparrow$} &\multicolumn{3}{c}{Correct \& Sec@$k\uparrow$} \\
    \cmidrule{2-10}
    & $k=1$ & $k=10$ & $k=50$ & $k=1$ & $k=10$ & $k=50$ & $k=1$ & $k=10$ & $k=50$ \\
    \cmidrule{2-10}        
    \OURS          & 31.2 & 49.6 & 59.6 & 12.2 & 28.4 & 36.5 & 7.4 & 18.7 & 23.1 \\
    \rowcolor{gray!30} + human & \textbf{33.2$_{+2.0}$} & \textbf{53.9$_{+4.3}$} & \textbf{65.4$_{+5.8}$} & \textbf{13.5$_{+1.3}$} & \textbf{33.8$_{+5.4}$} & \textbf{42.3$_{+5.8}$} & \textbf{8.9$_{+1.5}$} & \textbf{21.9$_{+3.2}$} & \textbf{28.9$_{+5.8}$}\\
     \bottomrule
    \end{tabular}
\end{table*}
We conduct ablation studies to evaluate the effectiveness of each design component in \OURS.
Specifically, we introduce the following variants by incrementally adding the key components (in both training and inference):
(1) \emph{Finetuned}: neither functional specification nor editing rule learning task is included;
(2) \emph{func-spec}: only the functional specification learning task is included;
(3) \emph{edit-rule}: only the editing rule learning task is included.
 
As shown in \cref{tab:ablation}, adding the functional specification learning task alone outperforms the end-to-end finetuning baselines on the functional correctness metrics across all three tasks by up to 21.8\%, while also improving the editing performance, e.g., efficiency, readability, and security, by up to 26.7\%.
Similarly, training models with the editing rule learning task alone can outperform the end-to-end finetuning baselines on the editing performance metrics across all three tasks by up to 33.2\%, while improving the functional correctness by up to 17.1\%. 
When combined, \OURS can outperform the end-to-end finetuning baselines by up to 34.8\% in overall editing performance.

While \OURS underperforms these ablation variants on some specific metrics, i.e., \emph{Correct@8} for performance, \emph{Compilability} for decompilation, and \emph{security@50} for security hardening, \OURS obtains the best performance when considering both the functional correctness and editing performance.

\begin{figure}[!t]
    \centering
    
    \includegraphics[width=0.6\linewidth]{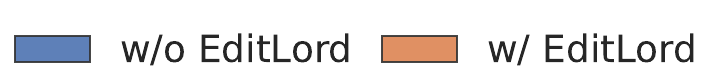}
    
    \subfloat[Performance opt.]{
        \includegraphics[width=0.31\linewidth]{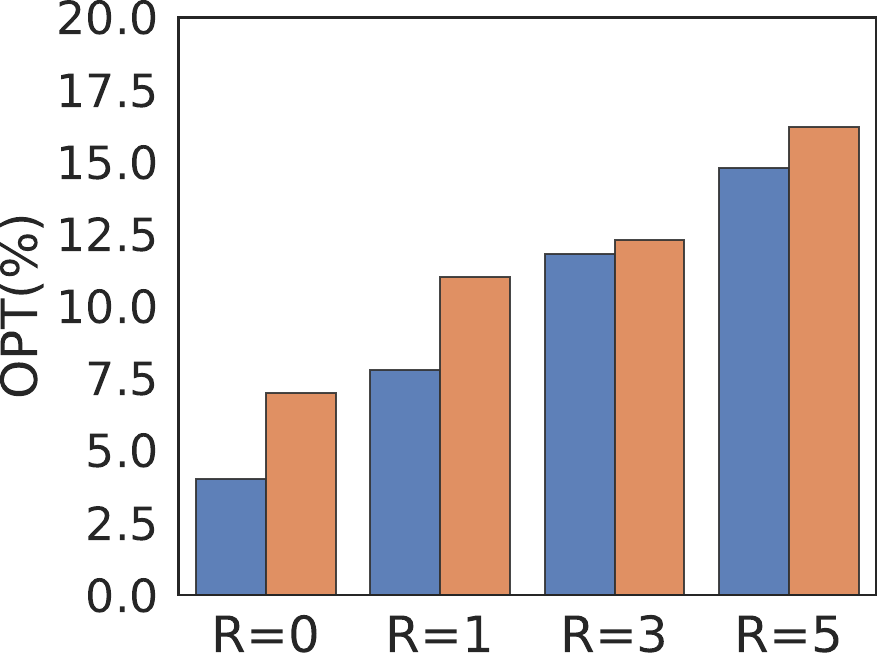}
    }
    \subfloat[Decompilation]{
        \includegraphics[width=0.3\linewidth]{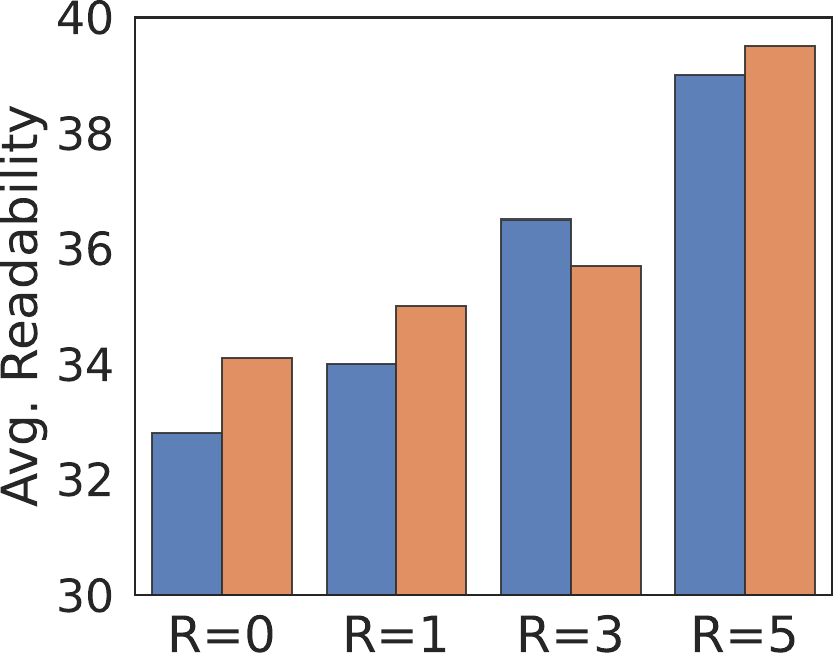}
    }    
    \subfloat[Security hardening]{
        \includegraphics[width=0.31\linewidth]{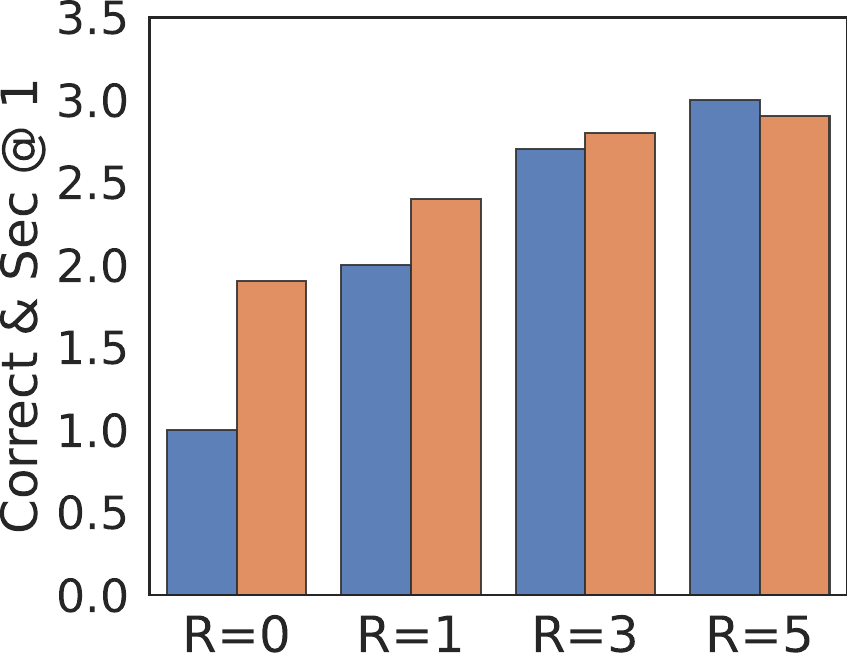}
    }
    
    \caption{\OURS with different editing modes under zero-shot prompting ($R=0$) and iterative refinement ($R=1,3,5$) with execution feedback. Here, $R$ is the number of iterations taken to refine the edited code.}
    \label{fig:editing-paradigm}
\end{figure}

\subsection{Other Editing Modes}
\label{subsec:eval_other_editing_modes}
In addition to finetuning, \OURS is complementary to other code editing modes.
We evaluate the effectiveness of \OURS when applied to zero-shot prompting and iterative refinement with execution feedback~\cite{peng2024perfcodegen, huang2024effi}.

As shown in \cref{fig:editing-paradigm}, \OURS improves the performance of zero-shot prompting and iterative refinement with execution feedback by an average of 56.3\% and 5.7\%, respectively.
We observe \OURS underperforms slightly in the decompilation and security hardening task when $R=3$ or $R=5$.
We suspect this is due to the coarse-grained execution feedback, e.g., compilation error, execution error, etc., while the state-of-the-art baselines incorporate more fine-grained information, e.g., per-statement execution profile~\cite{huang2024effi} (see \cref{sec:discussion} for the discussion). 

\subsection{Robustness and Generalization}
\label{subsec:eval_generalization_robustness}

\paragraph{Robustness}
We compare \OURS to the state-of-the-art baseline against semantics-preserving code transformations~\cite{wang2022recode, yefet2020adversarial, gao2023discrete, yang2022natural, bielik2020adversarial}. 
The transformations include renaming variables and functions to long random strings and removing comments, whitespaces, unused variables, and unused headers.
We observe individual transformations alone rarely produce significant output changes.
We thus exhaust each transformation (e.g., renaming all variables) and compose all of these transformations together to introduce input changes substantial enough.

\cref{tab:robustness-decompile-obf} (left) shows the average performance degradation in functional correctness, comparability, and readability when the input undergoes semantics-preserving transformations.
\OURS achieves up to 58.1\% less drops in readability measured by character-level readability compared to the baseline, indicating that producing explicit functional specifications and editing rules helps \OURS stay less susceptible to syntactic code changes.

\paragraph{Length generalization}
We investigate \OURS's generalizability to longer sequences than those seen in training~\cite{anil2022exploring}. 
Specifically, we select 71 testing samples with lengths over 500 and finetune \OURS only on 6,105 strictly shorter ($<500$) training samples.

\cref{tab:generalization-decompile-length} (right) demonstrates that \OURS suffers reduced performance degradation on unseen longer code, achieving up to 25.86\% less drops in readability measured by embedding-level readability (\cref{subsec:eval_setup}).

\subsection{Rule Augmentation by Human Experts}
\label{subsec:eval_rule_augment}
As \OURS produces explicit editing steps, it facilitates human intervention to introduce customized editing rules.
To assess the effectiveness of \OURS augmented by human intervention, we incorporate two human experts (both are the authors of this paper) to refine \OURS's generated editing rules and functional specifications for each testing sample, e.g., removing unreasonable rules or appending effective new rules.

\cref{tab:human} shows that such an augmentation enhances \OURS with up to 15.6\%, 35.5\%, and 25.1\% improvement in efficiency, readability in decompilation, and security.
\section{Discussion and Limitation}
\label{sec:discussion}
\paragraph{Iterative refinement with execution feedback}
Existing LM-based code editing approaches often leverage iterative refinement with execution feedback~\cite{huang2024effi, peng2024perfcodegen, xia2024automated, waghjale2024ecco}, which relies on the availability of test inputs.
However, the code to be edited may not always be well-maintained.
Therefore, in this paper, we do not assume the tests are available by default (\cref{sec:eval}).
We also show that \OURS is complementary to iterative refinement with coarse-grained execution feedback (\cref{subsec:method_editing_paradigm}).
We aim to study whether the fine-grained execution feedback, i.e., per-statement execution profile~\cite{huang2024effi}, can further improve \OURS in the future.

\vspace{.1cm}\noindent\textbf{Functional correctness guarantee~}
In code editing, functional correctness is arguably a hard constraint that cannot be violated, while the editing goals can be soft objectives.
While we ensured that our editing performance is measured strictly on the subset of edited code that must first be functionally correct, \OURS cannot \emph{guarantee} all the edited code is correct.
While this issue can be trivially mitigated by falling back to the original code, it completely fails to introduce any edits.
We thus follow the existing approaches~\cite{pieperf,tan2024llm4decompile} by treating functional correctness as a soft constraint.
Imposing a formal correctness guarantee for the edited code would be extremely valuable for future work.
\section{Related Work}

Code LMs have been extensively used to assist developers in editing code~\cite{guo2024codeeditorbench, li2023codeeditor, labash2024res, li2023instructcoder, cassano2023can, chakraborty2020codit, liu2024coedpilot, gupta2023grace, muennighoff2023octopack, singhal2024nofuneval} to fulfill various goals, such as bug fixing~\cite{xia2024automated, fan2023automated, liu2024marscode}, decompilation~\cite{tan2024llm4decompile, wong2023refining, xie2024resym, hu2024degpt}, efficiency optimization~\cite{huang2024effi, pieperf, garg2022deepperf, peng2024perfcodegen}, vulnerability repair~\cite{he2023sven,peng2025cweval,fu2022vulrepair,
xia2024automated, perry2023users, bhatt2023purple}, code translation~\cite{pan2024lost, eniser2024towards}, code refactoring~\cite{shirafuji2023refactoring, cummins2024don}, and more broadly related tasks like generating proofs~\cite{chen2024automated, chakraborty2024towards}, invariants~\cite{kamath2023finding,chakraborty2023ranking}, or specifications~\cite{murphy2024combining, ma2024specgen}.

Most existing approaches either adopt finetuning to learn the direct mapping between the code pairs or leverage iterative refinement with execution or self-generated feedbacks~\cite{huang2024effi, peng2024perfcodegen, huang2023agentcoder, xia2024automated, chen2023teaching, dong2024self, madaan2024self, zelikman2023self,liu2024learning}.
In contrast, \OURS complements these approaches by making the intermediate editing steps explicit.

\OURS shares a similar philosophy to bootstrapping the symbolic reasoning of LMs~\cite{zelikman2022star, kim2023language, huang2022large, chen2023teaching, hsieh2023distilling, wang2022self, zhang2022automatic, lightman2023let, zhou2024self}, where the LMs synthesize the reasoning procedures to self-augment the samples for supervised finetuning or preference tuning.
\OURS extends the idea by distilling a concise and composable editing meta-rule set, such that the code to be edited can share reusable editing steps and benefits from the improved generalization (\cref{subsec:eval_generalization_robustness}).
\section{Conclusion}
We introduced \OURS, a generic code editing framework, by learning the inductive code transformation rules to elicit the explicit code editing steps. 
Our key approach is to employ a language model (LM) as an inductive learner to distill a concise and composable meta-rule set for code editing from the training code pairs.
\OURS substantially outperforms the state-of-the-art code editing techniques in editing performance while enjoying significantly higher functional correctness and improved robustness against semantics-preserving code transformations across multiple critical software engineering and security applications, LM models, and editing modes.

\newpage
\section*{Impact Statement}
Code editing plays a crucial role in assisting developers' daily jobs. 
While Large language models (LLMs) have demonstrated promising capabilities in automated code transformation, their end-to-end nature often leads to hallucinated edits.
These hallucinations would be especially harmful when LLMs are used to edit security and safety-critical code.
Our paper introduced a new approach to improving the safety of LLMs when applied to security-critical software engineering applications.


\bibliography{reference}
\bibliographystyle{icml2025}

\newpage
\appendix
\onecolumn

\section{Prompt Format}
\subsection{Performance Optimization}
\label{app_sec:prompting-perf}

\begin{PromptBox}{Baseline prompt for performance optimization.}
\label{prompt:perf-base}
\textbf{User:} This is the slow code:\\
\text{[SLOW CODE]}\\
\{src\_code\}\\
\text{[/SLOW CODE]}\\
The corresponding fast code is:\\
\text{[FAST CODE]}\\
\textbf{Assistant:}\{tgt\_code\}\\
\text{[/FAST CODE]}
\end{PromptBox}

\begin{PromptBox}{Functional specifications only prompt for performance optimization.}
\label{prompt:perf-func-spec}
\textbf{User:} This is the slow code:\\
\text{[SLOW CODE]}\\
\{src\_code\}\\
\text{[/SLOW CODE]}\\
The given code describes the following problem: \{functional\_specification\}\\
The input specification is: \{input\_specification\}\\
The output specification is: \{output\_specification\}\\
The corresponding fast code is:\\
\text{[FAST CODE]}\\
\textbf{Assistant:}\{tgt\_code\}\\
\text{[/FAST CODE]}
\end{PromptBox}

\begin{PromptBox}{Editing rules only prompt for performance optimization.}
\label{prompt:perf-edit-spec}
\textbf{User:} This is the slow code:\\
\text{[SLOW CODE]}\\
\{src\_code\}\\
\text{[/SLOW CODE]}\\
Following editing rules should be applied: \{editing\_rules\}\\
The corresponding fast code is:\\
\text{[FAST CODE]}\\
\textbf{Assistant:}\{tgt\_code\}\\
\text{[/FAST CODE]}
\end{PromptBox}

\begin{PromptBox}{Our prompt for performance optimization.}
\label{prompt:perf-our}
\textbf{User:} This is the slow code:\\
\text{[SLOW CODE]}\\
\{src\_code\}\\
\text{[/SLOW CODE]}\\
The given code describes the following problem: \{functional\_specification\}\\
The input specification is: \{input\_specification\}\\
The output specification is: \{output\_specification\}\\
Following editing rules should be applied: \{editing\_rules\}\\
The corresponding fast code is:\\
\text{[FAST CODE]}\\
\textbf{Assistant:}\{tgt\_code\}\\
\text{[/FAST CODE]}
\end{PromptBox}

\subsection{Decompilation}
\label{app_sec:prompting-decompile}
\begin{PromptBox}{Baseline prompt for decompilation.}
\label{prompt:decompile-base}
\textbf{User:} This is the decompiled code:\\
\text{[MACHINE DECOMPILED CODE]}\\
\{src\_code\}\\
\text{[/MACHINE DECOMPILED CODE]}\\
The corresponding source code is:\\
\text{[ORIGINAL SOURCE CODE]}\\
\textbf{Assistant:}\{tgt\_code\}\\
\text{[/ORIGINAL SOURCE CODE]}
\end{PromptBox}

\begin{PromptBox}{Functional specifications only prompt for decompilation optimization.}
\label{prompt:decompile-func-spec}
\textbf{User:} This is the decompiled code:\\
\text{[MACHINE DECOMPILED CODE]}\\
\{src\_code\}\\
\text{[/MACHINE DECOMPILED CODE]}\\
The given code describes the following problem: \{functional\_specification\}\\
The input specification is: \{input\_specification\}\\
The output specification is: \{output\_specification\}\\
The corresponding source code is:\\
\text{[ORIGINAL SOURCE CODE]}\\
\textbf{Assistant:}\{tgt\_code\}\\
\text{[/ORIGINAL SOURCE CODE]}
\end{PromptBox}

\begin{PromptBox}{Editing rules only prompt for performance decompilation.}
\label{prompt:decompile-edit-spec}
\textbf{User:} This is the decompiled code:\\
\text{[MACHINE DECOMPILED CODE]}\\
\{src\_code\}\\
\text{[/MACHINE DECOMPILED CODE]}\\
Following editing rules should be applied: \{editing\_rules\}\\
The corresponding source code is:\\
\text{[ORIGINAL SOURCE CODE]}\\
\textbf{Assistant:}\{tgt\_code\}\\
\text{[/ORIGINAL SOURCE CODE]}
\end{PromptBox}

\begin{PromptBox}{Our prompt for decompilation.}
\label{prompt:decompile-our}
\textbf{User:} This is the decompiled code:\\
\text{[MACHINE DECOMPILED CODE]}\\
\{src\_code\}\\
\text{[/MACHINE DECOMPILED CODE]}\\
The given code describes the following problem: \{functional\_specification\}\\
The input specification is: \{input\_specification\}\\
The output specification is: \{output\_specification\}\\
Following editing rules should be applied: \{editing\_rules\}\\
The corresponding source code is:\\
\text{[ORIGINAL SOURCE CODE]}\\
\textbf{Assistant:}\{tgt\_code\}\\
\text{[/ORIGINAL SOURCE CODE]}
\end{PromptBox}

\subsection{Security}
\label{app_sec:prompting-sec}
\begin{PromptBox}{Baseline prompt for security.}
\label{prompt:security-base}
\textbf{User:} This is the vulnerable code:\\
\text{[VULNERABLE CODE]}\\
\{src\_code\}\\
\text{[/VULNERABLE CODE]}\\
The corresponding secure code is:\\
\text{[SECURE CODE]}\\
\textbf{Assistant:}\{tgt\_code\}\\
\text{[/SECURE CODE]}
\end{PromptBox}

\begin{PromptBox}{Functional specifications only prompt for security.}
\label{prompt:security-func-spec}
\textbf{User:} This is the vulnerable code:\\
\text{[VULNERABLE CODE]}\\
\{src\_code\}\\
\text{[/VULNERABLE CODE]}\\
The given code describes the following problem: \{functional\_specification\}\\
The input specification is: \{input\_specification\}\\
The output specification is: \{output\_specification\}\\
The corresponding secure code is:\\
\text{[SECURE CODE]}\\
\textbf{Assistant:}\{tgt\_code\}\\
\text{[/SECURE CODE]}
\end{PromptBox}

\begin{PromptBox}{Editing rules only prompt for security.}\label{prompt:security-edit-spec}
\textbf{User:} This is the vulnerable code:\\
\text{[VULNERABLE CODE]}\\
\{src\_code\}\\
\text{[/VULNERABLE CODE]}\\
Following editing rules should be applied: \{editing\_rules\}\\
The corresponding secure code is:\\
\text{[SECURE CODE]}\\
\textbf{Assistant:}\{tgt\_code\}\\
\text{[/SECURE CODE]}
\end{PromptBox}

\begin{PromptBox}{Our prompt for security.}
\label{prompt:security-our}
\textbf{User:} This is the vulnerable code:\\
\text{[VULNERABLE CODE]}\\
\{src\_code\}\\
\text{[/VULNERABLE CODE]}\\
The given code describes the following problem: \{functional\_specification\}\\
The input specification is: \{input\_specification\}\\
The output specification is: \{output\_specification\}\\
Following editing rules should be applied: \{editing\_rules\}\\
The corresponding secure code is:\\
\text{[SECURE CODE]}\\
\textbf{Assistant:}\{tgt\_code\}\\
\text{[/SECURE CODE]}
\end{PromptBox}

\subsection{Meta-Rule Set}
\label{app_sec:meta-rule}
\begin{PromptBox}{Generic/specific evaluation for an editing rule}
Please analyze the provided editing rule (in order to improve \{task\_name\}) and determine whether it is broadly applicable across different code snippets (generic) or tailored to a specific code snippet (specific). An editing rule like "\{generic\_rule\_example\}" should be considered as a generic rule. While a rule like "\{specific\_rule\_example\}" should be considered as a specific rule.

Provide your response in the following format:

The rule is [generic/specific] because ...

So, what do you think about the rule "\{editing\_rule\}"? Is it generic or specific?
\end{PromptBox}

\begin{PromptBox}{Add/Merge rules}
Please analyze the provided editing rule (in order to improve \{task\_name\}) and compare it with the existing editing rules in the meta-rule set. If it's similar to any existing editing rule, please suggest how it should be integrated into the existing meta-rule set. Specify the one and only one appropriate action from the options below:

[ADD]: If none of the existing editing rules in the meta-rule set is similar to the current one, provide the refined and updated editing rule to be added to the set.

[MERGE]: If the current editing rule is similar to an existing editing rule, indicate which existing meta-rule is similar to the current editing rule so that they can be merged and how they should be merged.

If [ADD] is selected, please provide the refined and updated editing rule to be added to the set directly without any other information. If [MERGE] is selected, please provide exactly the existing meta-rule that is similar to the current editing rule with an updated editing rule.

Please notice that whether you select [ADD] or [MERGE], the editing rule you add or merge into the meta-rule set must adhere to the format ``switch from ... to ...''. Ensure that you only provide editing rules that transition from a \{old\_property\} to \{new\_property\}.

Here are several examples of the output:

[Example Output 1]

[ADD] only the editing rule to be added here [/ADD]

[/Example Output 1]

[Example Output 2]

[MERGE] only the editing rule to be merged and the updated rule, split by semicolon [/MERGE]

[/Example Output 2]

Meta-Rule Set:

\{meta\_rule\_set\}

Editing Rule Requested for Analysis:

\{editing\_rule\}
\end{PromptBox}

\section{Meta-Rule Set Learned by \OURS}
\cref{tab:rule-example} shows some examples of meta-rules for code editing discovered by \OURS. 
We also calculate the percentage of edited samples (as indicated in the third column) that benefit from the specific rule in each row.
For example, the rule ``switch from \texttt{cout} to \texttt{printf}'' applies to 32.4\% of the testing samples that have obtained performance improvement.
\begin{table}[!t]

    \centering
    \setlength{\tabcolsep}{3pt}
    \renewcommand{\arraystretch}{1.1}
    \small
    
    \caption{Meta-rule examples discovered by \OURS for performance optimization, decompilation, and security hardening (\cref{subsec:method_rule_learning}). Each rule follows the universal format ``switch \emph{from} \texttt{[old properties]} \emph{to} \texttt{[new properties]}''. The third column shows the percentage of the samples improved by the specific rule.}
    \label{tab:rule-example}
    
    \begin{tabular}{llrl}
    \toprule
    From & To & \%$\uparrow$ & Examples \\
    \midrule
    \multicolumn{3}{l}{\bf Performance Optimization} \\
    \hdashline[3pt/5pt]
    \rowcolor{gray!30} \texttt{cout} & \texttt{printf} & 32.4 & \cref{fig:perf-case-3}\\
    \texttt{cin} & \texttt{scanf} & 24.8 & \cref{fig:perf-case-3}\\
    \rowcolor{gray!30} multiple nested loops for condition checks & a single streamlined loop for condition checks and assignments & 21.0 & \cref{fig:perf-case-1} \\
    recursive function calls & optimized iterative data handling methods & 7.6 & \cref{fig:perf-case-1} \\
    \rowcolor{gray!30} dynamic memory allocation & static memory allocation & 7.6 & \cref{fig:perf-case-3}\\
    \midrule
    \multicolumn{3}{l}{\bf Decompilation} \\
    \hdashline[3pt/5pt] 
    \rowcolor{gray!30} complex if-else structure & simplified conditional logic & 11.9  & \cref{fig:decompile-case-2} \\
    complex pointer arithmetic & clear variable assignments & 10.1 & \cref{fig:decompile-case-2}, \ref{fig:decompile-case-1}\\
    \rowcolor{gray!30} redundant checks & straightforward boolean comparisons & 6.0 & \cref{fig:decompile-case-1}\\
    cryptic variable names & descriptive variable names & 5.9 & \cref{fig:decompile-case-2}\\
    \rowcolor{gray!30} ambiguous function signatures & clear function signatures & 5.9 & \cref{fig:decompile-case-1} \\
    \midrule
    \multicolumn{3}{l}{\bf Security Hardening}\\
    \hdashline[3pt/5pt] 
    \rowcolor{gray!30} no checks on function return values & check function return value & 32.9 & \cref{fig:secure-case-1}\\
    direct SQL string interpolation & use of parameterized logic & 10.6 & \cref{fig:secure-case-4}\\
    \rowcolor{gray!30} unvalidated input handling & check for buffer overflows on memory accesses & 8.9 & \cref{fig:secure-case-2}\\
    unvalidated memory allocation & check maximum buffer allocation size before allocation & 7.6 & \cref{fig:secure-case-3}\\
    \rowcolor{gray!30} direct parsing & implement comprehensive validation checks for character handling & 6.8 & \cref{fig:secure-case-5}\\
    \bottomrule
    \end{tabular}
\end{table}
\FloatBarrier

\section{Case Study}
\subsection{Performance Optimization}
We show several examples to demonstrate how \OURS improve performance optimization.
\begin{figure}[ht]
    \centering
    \setlength{\tabcolsep}{5pt}
    \renewcommand{\arraystretch}{0}
    \begin{tabular}[!t]{ p{7.4cm} p{9cm}}
    Given slow code & \OURS output \\
    \begin{mdframed}
        \input{case-study/perf-src-code-1}
    \end{mdframed} 
    & 
    \begin{mdframed}
        \input{case-study/perf-gen-code-1}
    \end{mdframed}
    \end{tabular}
    \caption{Performance optimization example 1.}
    \label{fig:perf-case-1}
\end{figure}

\begin{figure}[!t]
    \centering
    \setlength{\tabcolsep}{5pt}
    \renewcommand{\arraystretch}{0}
    \begin{tabular}[!t]{ p{7.4cm} p{9cm}}
    Given slow code & \OURS output \\
    \begin{mdframed}
        \input{case-study/perf-src-code-3}
    \end{mdframed} 
    & 
    \begin{mdframed}
        \input{case-study/perf-gen-code-3}
    \end{mdframed}
    \end{tabular} 
    \caption{Performance optimization example 2.}
    \label{fig:perf-case-3}
\end{figure}

\FloatBarrier

\subsection{Decompilation}
We show several examples to demonstrate how \OURS improves the readability of machine-decompiled code. 
\begin{figure}[ht]
    \centering    
    \setlength{\tabcolsep}{5pt}
    \renewcommand{\arraystretch}{0}
    \begin{tabular}[!t]{ p{7.4cm} p{9cm}}
    Given machine decompiled code & \OURS output \\
    \begin{mdframed}
        \input{case-study/decompile-src-code-2}
    \end{mdframed} 
    & 
    \begin{mdframed}
        \input{case-study/decompile-gen-code-2}
    \end{mdframed}
    \end{tabular} 
    \caption{Decomilation example 1.}
    \label{fig:decompile-case-2}
\end{figure}

\begin{figure}[!t]
    \centering    
    \setlength{\tabcolsep}{5pt}
    \renewcommand{\arraystretch}{0}
    \begin{tabular}[!t]{ p{7.4cm} p{9cm}}
    Given machine decompiled code & \OURS output \\
    \begin{mdframed}
        \input{case-study/decompile-src-code-1}
    \end{mdframed} 
    & 
    \begin{mdframed}
        \input{case-study/decompile-gen-code-1}
    \end{mdframed}
    \end{tabular} 
    \caption{Decomilation example 2.}
    \label{fig:decompile-case-1}
\end{figure}

\FloatBarrier

\subsection{Security Hardening}
We show several examples to demonstrate how \OURS hardens the vulnerable code. 
\begin{figure}[ht]
    \centering    
    \setlength{\tabcolsep}{5pt}
    \renewcommand{\arraystretch}{0}
    \begin{tabular}[!t]{ p{7.4cm} p{9cm}}
    Given vulnerable code & \OURS output \\
    \begin{mdframed}
        \input{case-study/secure-src-code-1}
    \end{mdframed} 
    & 
    \begin{mdframed}
        \input{case-study/secure-gen-code-1}
    \end{mdframed}
    \end{tabular} 
    \caption{Secure hardening example 1.}
    \label{fig:secure-case-1}
\end{figure}

\begin{figure}[!t]
    \centering    
    \setlength{\tabcolsep}{5pt}
    \renewcommand{\arraystretch}{0}
    \begin{tabular}[!t]{ p{7.4cm} p{9cm}}
        Given vulnerable code & \OURS output \\
        \begin{mdframed}
            \input{case-study/secure-src-code-4}
        \end{mdframed} 
        & 
        \begin{mdframed}
            \input{case-study/secure-gen-code-4}
        \end{mdframed}
    \end{tabular} 
    \caption{Secure hardening example 2.}
    \label{fig:secure-case-4}
\end{figure}

\begin{figure}[!t]
    \centering    
    \setlength{\tabcolsep}{5pt}
    \renewcommand{\arraystretch}{0}
    \begin{tabular}[!t]{ p{7.4cm} p{9cm}}
        Given vulnerable code & \OURS output \\    
        \begin{mdframed}
            \input{case-study/secure-src-code-2}
        \end{mdframed} 
        & 
        \begin{mdframed}
            \input{case-study/secure-gen-code-2}
        \end{mdframed}
    \end{tabular} 
    \caption{Secure hardening example 3.}
    \label{fig:secure-case-2}
\end{figure}

\begin{figure}[!t]
    \centering    
    \setlength{\tabcolsep}{5pt}
    \renewcommand{\arraystretch}{0}
    \begin{tabular}[!t]{ p{7.4cm} p{9cm}}
        Given vulnerable code & \OURS output \\
        \begin{mdframed}
            \input{case-study/secure-src-code-3}
        \end{mdframed} 
        & 
        \begin{mdframed}
            \input{case-study/secure-gen-code-3}
        \end{mdframed}
    \end{tabular} 
    \caption{Secure hardening example 4.}
    \label{fig:secure-case-3}
\end{figure}

\begin{figure}[!t]
    \centering    
    \setlength{\tabcolsep}{5pt}
    \renewcommand{\arraystretch}{0}
    \begin{tabular}[!t]{ p{7.4cm} p{9cm}}
        Given vulnerable code & \OURS output \\
        \begin{mdframed}
            \input{case-study/secure-src-code-5}
        \end{mdframed} 
        & 
        \begin{mdframed}
            \input{case-study/secure-gen-code-5}
        \end{mdframed}
    \end{tabular} 
    \caption{Secure hardening example 5.}
    \label{fig:secure-case-5}
\end{figure}

\FloatBarrier
\section{Rule Learning Details}

Given the meta-rule set $G=\{r|r=\text{pre construct}\rightarrow\text{post construct}\}$ (\cref{fig:methodology}), we can manifest these meta-rules to each training sample for augmentation. 
However, directly prompting the LM to infer meta-rules for each pre-edit and post-edit code pair $(x_i,y_i)$ often leads to hallucinations. 
To mitigate this, we propose a two-step approach. 
First, we apply the LM to extract pre construct rules from $x_i$ and post construct rules from $y_i$ separately. 
Then, we identify which of these pre/post construct combinations are included in the meta rule set. 
We adopt these verified combined rules as the meta-rules for the code pair $(x_i, y_i)$.

\section{Hyperparameters}
To finetune DeepSeek-Coder, we use a default batch size of 32, a learning rate of 1e-5, and 4,000 context lengths for both the input and output tokens. 
The models are optimized using AdamW and trained for a fixed number of 10 epochs, and we use the model checkpoint that achieves the best validation loss for inference. 
To finetune GPT-4o mini, we train for only one epoch. 
At the inference stage, we set the temperature to 0.7 and use the model's default window size, i.e., 16K for DeepSeek-Coder and 128K for GPT-4o-mini.

\section{Additional Experiments}
\subsection{Extra Well-Known Benchmark}
\begin{table}[!t]
    \centering
    \setlength{\tabcolsep}{2.8pt}
    \renewcommand{\arraystretch}{1.1}
    \small
    \caption{Evaluating DeepSeek-Coder 1.3B on the Code Polish task in CodeEditorBench.}
    \label{tab:codeeditorbench}
    \begin{tabular}{rlll}
    \toprule
    & Accuracy & OptScoreTime & OptScore \\
     \midrule
     Finetuned & 0.9\% & 0.03\% & 0.09\%\\
     \rowcolor{gray!30} \OURS (Ours)  & \textbf{23.4\%} & \textbf{1.83\%} & \textbf{1.19\%}\\ 
    \bottomrule 
    \end{tabular}
\end{table}

We further evaluated our finetuned DeepSeek-Coder 1.3B on the Code Polish task in CodeEditorBench~\citep{guo2024codeeditorbench}. 
We follow their metrics by focusing on 1) accuracy: the percentage of problems with correct edits; 2) OptScoreTime: the execution time improvement; and 3) OptScore, the improvement computed by the averaged time and memory.
As illustrated in \cref{tab:codeeditorbench}, \OURS, even without extra finetuning on this dataset, outperforms the baseline by 22.5\%, 1.8\%, and 1.1\%, respectively.

\subsection{Out-of-Domain Generalization}
\begin{table}[!t]
    \centering
    \setlength{\tabcolsep}{2.8pt}
    \renewcommand{\arraystretch}{1.1}
    \small
    
    \caption{Unseen CWEs evaluation. We also include evaluation on seen CWEs.}
    
    \label{tab:robustness-unseen-cwes}
    \begin{tabular}{rrlllllllll}
    \toprule
    & &\multicolumn{3}{c}{Correct@$k\uparrow$} &\multicolumn{3}{c}{Security@$k\uparrow$} &\multicolumn{3}{c}{Correct \& Sec@$k\uparrow$} \\
        \cmidrule{2-10}
        & $k=1$ & $k=10$ & $k=50$ & $k=1$ & $k=10$ & $k=50$ & $k=1$ & $k=10$ & $k=50$ \\
        \midrule
    \multirow{2}{*}{Seen CWEs} & Finetuned & 24.3 & 38.4	& 41.7 & \textbf{12.8} & \textbf{44.0} & 50.0 & 7.7 & 21.6 & \textbf{25.0}\\
        & \cellcolor{gray!30}\OURS (Ours) & \cellcolor{gray!30}\textbf{36.8} & \cellcolor{gray!30}\textbf{53.3} & \cellcolor{gray!30}\textbf{66.7} & \cellcolor{gray!30}12.5 & \cellcolor{gray!30}43.3 & \cellcolor{gray!30}\textbf{58.3} & \cellcolor{gray!30}\textbf{8.7} & \cellcolor{gray!30}\textbf{24.7} & \cellcolor{gray!30}\textbf{25.0} \\  
        \multirow{2}{*}{Unseen CWEs} & Finetuned & 24.1 & 35.0 & 40.0 & 8.6 & 14.8 & 22.5 & 4.6 & 11.5 & 17.5 \\
        & \cellcolor{gray!30}\OURS (Ours) & \cellcolor{gray!30}\textbf{29.6} & \cellcolor{gray!30}\textbf{48.5} & \cellcolor{gray!30}\textbf{57.5} & \cellcolor{gray!30}\textbf{12.1} & \cellcolor{gray!30}\textbf{23.8} & \cellcolor{gray!30}\textbf{30.0} & \cellcolor{gray!30}\textbf{7.0} & \cellcolor{gray!30}\textbf{16.9} & \cellcolor{gray!30}\textbf{22.5}\\
    \bottomrule
        
    \end{tabular}
\end{table}

\begin{table}[!t]
    \centering
    \setlength{\tabcolsep}{2.8pt}
    \renewcommand{\arraystretch}{1.1}
    \small
    
    \caption{Unseen programming languages evaluation. We also include evaluation on seen language.}
    
    \label{tab:robustness-unseen-lang}
    \begin{tabular}{rrlll}
    \toprule
    & & Accuracy & OptScoreTime & OptScore \\
     \midrule
    \multirow{2}{*}{Seen Language (cpp)} & Finetuned & 1.4\% & 0.02\% & 0.24\% \\
        & \cellcolor{gray!30}\OURS (Ours) & \cellcolor{gray!30}\textbf{28.3\%} & \cellcolor{gray!30}\textbf{3.1\%} & \cellcolor{gray!30}\textbf{2.29\%} \\  
        \multirow{2}{*}{Unseen Languages} & Finetuned & 0.7\% & 0.04\% & 0.02\% \\
        & \cellcolor{gray!30}\OURS (Ours) & \cellcolor{gray!30}\textbf{20.9\%} & \cellcolor{gray!30}\textbf{1.18\%} & \cellcolor{gray!30}\textbf{0.63\%} \\  
    \bottomrule
        
    \end{tabular}
\end{table}

\paragraph{Unseen CWEs.}
As described in \cref{subsec:eval_setup}, our training comes from SVEN~\citep{he2023sven}, but our testing is from CWEval~\citep{peng2025cweval} with unseen CWEs. 
We further analyze the performance of the baseline and \OURS on unseen CWEs.
As shown in \cref{tab:robustness-unseen-cwes}, \OURS generalizes better on unseen CWEs, outperforming the baseline by 38.1\%. 

\paragraph{Unseen languages.}
We also investigate \OURS's generalizability to unseen languages (Python/Java) in performance optimization tasks in CodeEditorBench~\citep{guo2024codeeditorbench} when training on C++ code only.
Specifically, we train on the HQ dataset from \citet{pieperf}, which contains only C++ samples.
\cref{tab:robustness-unseen-lang} demonstrates that \OURS maintains strong generalization to unseen languages, outperforming the baseline by 20.2\% in accuracy, 1.14\% in execution time improvement, and 0.61\% in combined time and memory efficiency improvement.

\subsection{Data Efficiency}
\begin{table}[!t]

    \centering
    \setlength{\tabcolsep}{3pt}
    \renewcommand{\arraystretch}{1.1}
    \small
    
    \caption{Comparing \OURS to the finetuned baseline under varying amounts of training data.}
    \label{tab:data-efficiency}
    \begin{tabular}{rrlllll}
    \toprule
    & \multirow{2}{*}{Data Usage} & \multirow{2}{*}{Compile$_\uparrow$} & \multirow{2}{*}{Correct$_\uparrow$} & \multicolumn{3}{c}{Readability$_\uparrow$} \\
    \cmidrule{5-7}
    & & & & char & token & emb \\
    \midrule 
    \multicolumn{6}{l}{Robustness}\\
    \midrule
    Finetuned & 50\% & 38.9 & 77.1 & 36.6 & 40.8 & 37.5 \\
    \rowcolor{gray!30} & 50\% & 41.2 & \textbf{93.1} & 42.6 & 46.3 & 41.4 \\
    \rowcolor{gray!30} \multirow{-2}{*}{\OURS (Ours)}& 100\% & \textbf{46.6} & \textbf{93.1} & \textbf{44.0} & \textbf{47.6} & \textbf{41.4} \\

    \bottomrule
    \end{tabular}
\end{table}

To evaluate how \OURS scale with varying amounts of finetuning data, we conduct an additional experiment on the decompilation task using only 50\% of the finetuning dataset. 
As shown in \cref{tab:data-efficiency}, \OURS, trained with just 50\% of the data, still surpasses the baseline trained on the full dataset by 5.9\%. 
Moreover, \OURS achieves significantly better readability, outperforming the baseline by 13.5\% and 16.4\% on character- and token-level readability metrics, respectively. 
This highlights the sample efficiency of \OURS, requiring less than 50\% of training samples while achieving comparable performance.



\end{document}